\documentclass[a4paper,12pt]{article}
\usepackage{graphicx}
\usepackage{amsmath}
\usepackage{amssymb}
\usepackage{a4wide}

\newcommand\pubnumber{FLAVOUR(267104)-ERC-83}
\newcommand\pubdate{\today}

\def\address{\
TUM Institute for Advanced Study,\\ Lichtenbergstra\ss e 2a, 85748 Garching, GERMANY}
\def\support{\footnote{Work supported by ERC Advanced Grant project ``FLAVOUR'' (267104).}}

\def\Title#1{\begin{center} {\Large #1 } \end{center}}
\def\Author#1{\begin{center}{ \sc #1} \end{center}}
\def\Address#1{\begin{center}{ \it #1} \end{center}}

\newcommand\pubblock{\rightline{\begin{tabular}{l} \pubnumber\\
         \pubdate  \end{tabular}}}
\newenvironment{Abstract}{\begin{quotation}}{\end{quotation}}
\newenvironment{Presented}{\begin{quotation}\begin{center} 
             PRESENTED AT\end{center}\bigskip 
      \begin{center}\begin{large}}{\end{large}\end{center} \end{quotation}}
\def\Acknowledgements{\bigskip  \bigskip \begin{center} \begin{large}
             \bf ACKNOWLEDGEMENTS \end{large}\end{center}}

\def\beq{\begin{equation}}
\def\eeq#1{\label{#1}\end{equation}}
\def\eeqn{\end{equation}}

\def\beqa{\begin{eqnarray}}
\def\eeqa#1{\label{#1}\end{eqnarray}}
\def\eeqan{\end{eqnarray}}

\let\bar=\overbar

\def\Dslash{\not{\hbox{\kern-4pt $D$}}}
\def\dslash{\not{\hbox{\kern-2pt $\del$}}}

\def\msb{{\bar{\ssstyle M \kern -1pt S}}}

\begin{document}
\begin{titlepage}
\pubblock

\vfill
\Title{Addressing Hadronic Uncertainties in Extractions of $\phi_s$}
\vfill
\Author{Robert Knegjens\support}
\Address{\address}
\vfill
\begin{Abstract}
In light of recent LHC results for the extraction of the $B_s$ mixing phase $\phi_s$, we can already conclude that if New Physics (NP) is present in this observable, it is hiding pretty well.
Thus, as our hunt continues, we must be weary not to confuse NP for penguin effects, or vice versa.
In this talk the progress made towards addressing hadronic uncertainties in extractions of $\phi_s$ from $B_s\to J/\psi \phi$ is reviewed,
and the nature of the scalar $f_0(980)$ state, which plays a dominant role in the extraction of $\phi_s$ from the $B_s\to J/\psi \pi^+\pi^-$ decay, is discussed.
\end{Abstract}
\vfill
\begin{Presented}
The 8th International Workshop on\\ the CKM Unitarity Triangle (CKM 2014),\\ Vienna, Austria, September 8-12, 2014
\end{Presented}
\vfill
\end{titlepage}
\def\thefootnote{\fnsymbol{footnote}}
\setcounter{footnote}{0}

\section{Introduction}

The CP-violating mixing phase $\phi_s$\footnote{We define $\phi_s$ as the argument of the dispersive part of $\langle B_s^0|{\cal H}|\overline{B}_s^0\rangle$, with phase conventions fixed so that $\phi^{\rm SM}_s = -2\beta_s$.} of the $B_s$ meson system has long been a promising and popular probe of New Physics (NP).
In principle, it can be extracted from its interference with the phase $\Delta\phi^f$ of any chosen decay mode $B_s\to f$ that is common to both $B_s$ flavour states,
provided, of course, that the latter {\it phase shift}\/ is known. 
The decay mode of choice is $B_s\to J/\psi \phi$, which has the desirable property that its dominant diagram has a vanishing phase shift relative to $\phi_s$,
as well as a favourable decay rate and experimental signature.
Yet it also has an Achilles heel: the contribution of penguin diagrams to the phase shift. 
Although suppressed, these diagrams introduce hadronic uncertainties that must be controlled to achieve precise measurements.

The vector-vector nature of the $J/\psi \phi$ final state gives three transversity (or polarization) amplitudes, two of which are CP-even and one CP-odd~\cite{Dighe:1995pd,Dighe:1998vk}, necessitating an angular analysis to disentangle them.
Because the $\phi$ is observed via its decay to $K^+ K^-$, a small S-wave component has also been found to contribute, adding a fourth, CP-odd, amplitude to the mix~\cite{Stone:2008ak,Xie:2009fs}.
The dominant resonance in this S-wave is the scalar $f_0(980)$ state~\cite{Aaij:2013orb}.
This led to a proposal 
to also extract $\phi_s$ from $B_s\to J/\psi \pi^+\pi^-$~\cite{Stone:2009hd}, which has a dominant $f_0(980)$ resonance, too~\cite{Aaij:2014emv}.
If the $f_0(980)$ can be shown to have a $s\bar s$ composition similar to the $\phi$, the phase shift associated with this decay mode could be controlled in an analogous way to those of $B_s\to J/\psi \phi$.

There has been great progress recently from the LHC experiments in extracting $\phi_s$.
For the decay mode $B_s\to J/\psi K^+ K^-$ the following values have been reported:
\begin{align}
    \phi_s\ (+\, \Delta\phi_h^{J/\psi KK}) =
    \left\{\begin{array}{rcll}
        -(3.3 \pm 2.8)^\circ &:&  {\rm LHCb}\ (3{\rm fb}^{-1}) &\cite{Aaij:2014zsa}\\
        \ (6.9 \pm 14.6)^\circ &:&  {\rm ATLAS}\ (5{\rm fb}^{-1})&\cite{Aad:2014cqa}\\
        -(1.7 \pm 6.5)^\circ &:&  {\rm CMS}\ (20{\rm fb}^{-1})&\cite{CMS-PAS-BPH-13-012}
    \end{array}\right..
     \label{BsKKresult}
\end{align}
Likewise, for the extraction from $B_s\to J/\psi \pi^+ \pi^-$ LHCb has reported~\cite{Aaij:2014dka}
\begin{align}
    \phi_s\ (+\, \Delta\phi^{J/\psi \pi\pi}) &= (4 \pm 3.9)^\circ \label{Bsf0result}
\end{align}
These results clearly lie close to the Standard Model (SM) prediction of $\phi_s^{\rm SM} = -(2.1\pm 0.1)^\circ$~\cite{Charles:2004jd}, dispelling the possibility of a large NP signal.
Thus if there is NP present in the $B_s$ mixing phase, it must be small and perhaps even be conspiring with the phase shifts to stay hidden.
Thus the dilemma is that without control over the phase shifts, it will be impossible to conclude whether a future experimental deviation indicates NP, or if no deviation means no NP.

In Section~\ref{sec:gold} we review the control of penguin contributions to the phase shifts $\Delta\phi^{J/\psi\phi}_h$.
In Section~\ref{sec:silver} we discuss whether the dominant resonance in $B_s\to J/\psi \pi^+\pi^-$, the scalar $f_0(980)$, can be interpreted as an $s\bar s$ state.
Finally in Section~\ref{sec:summary} we give a summary.

\section{Controlling penguin contributions in $B_s\to J/\psi \phi$}\label{sec:gold}

Let us denote the four transversity amplitudes of $B^0_s\to (J/\psi K^+ K^-)_h$ by $A_h$, and those of the CP-conjugate process by $\overline{A}_h$, with $h \in \{\parallel,\perp, 0, S\}$.
Although in practice the full angular analysis involves 20 observables, the mixing-induced CP-violating phases for each amplitude are in essence extracted from the tagged time-dependent analyses~\cite{Dighe:1998vk}:
\begin{equation}
    \frac{|A_h(t)|^2 - |\overline{A}_h(t)|^2}{|A_h(t)|^2 + |\overline{A}_h(t)|^2}
    = \frac{(1-|\lambda_h|^2) \cos(\Delta M_s t) + 2\,{\rm Im}(\lambda_h)\sin(\Delta M_s t)}{(1+|\lambda_h|^2) \cosh(\Delta \Gamma_s t) + 2\,{\rm Re}(\lambda_h)\sinh(\Delta \Gamma_s t)},
\end{equation}
with $\Delta M_s$ and $\Delta\Gamma_s$ the mass and decay width differences of the $B_s$ system, and
\begin{equation}
    \lambda_h 
    = - \eta_h \sqrt{\frac{1- C_h}{1+C_h}}\, e^{-i(\phi_s + \Delta\phi_h)},
    \quad \eta_h =
    \left\{\begin{array}{ccc}
        +1 &:&  h=\{\parallel, 0\}\\
        -1 &:&  h=\{\perp, S\}
    \end{array}\right..
\end{equation}
Here $C_h$ is the direct CP violation and $\Delta\phi_h$ the phase shift for each transversity amplitude.

A non-zero $\Delta\phi_h$ requires the sub-leading diagrams to carry CP-violating phases, which is notably the case for penguin diagrams with internal up and top quarks.
By pulling the CKM structure out of the contributing diagrams and utilizing the unitarity of the CKM matrix in the SM, we may write~\cite{Faller:2008gt}
\begin{equation}
A_h = A(B_s^0 \to (c\bar c\, \bar s s)_h) = {\cal A}_h\left[ 1 + \epsilon\, e^{i\gamma}\,b_h e^{i\theta_h}\right].
    \label{AmplSS}
\end{equation}
The ${\cal A}_h$ and $b_h e^{i\theta_h}$ are CP-conserving hadronic parameters with a non-perturbative nature.
In particular $b_h e^{i\theta_h}$ represents the ratio of such hadronic contributions from diagrams that carry a CP-violating phase ($\gamma$) with those that do not, whereby it is approximately a ratio of penguin contributions over the leading tree contribution.
A virtue of these decays is that $b_h$ is Cabibbo suppressed by $\epsilon \equiv \lambda^2/(1-\lambda^2)\simeq 5\%$.
Consequently, for $\gamma = 68^\circ$, our observables are\footnote{Note that also the CKM length $R_b\sim 0.4$ has been absorbed into the definition of $b_h$.}~\cite{Faller:2008gt}
\begin{align}
    \Delta\phi_h \approx (6^\circ)\cdot b_h\cos\theta_h,\qquad
    C_h \approx (-10\%)\cdot b_h\sin\theta_h. 
    \label{hadrObs}
\end{align}

The troublesome penguin diagrams are loop suppressed and also OZI suppressed due to the radiative production of the $J/\psi$, leading to a perturbative estimate of $b_h \sim {\cal O}(10^{-2})$~\cite{Boos:2004xp,Li:2006vq,Gronau:2008cc}.
The difficulty, however, is in accounting for non-perturbative hadronic effects, which could potentially give an order of magnitude enhancement. 
The presence or absence of such an enhancement makes all the difference in whether we can neglect the phase shift given in \eqref{hadrObs}, or not.

One method to control penguin contributions is via the approximate $SU(3)_{\rm F}$ flavour symmetry of strong interactions, which can relate the hadronic parameters of decay modes driven by $b\to s c\bar c$ transitions to those by $b\to d c\bar c$.
Analogous to \eqref{AmplSS}, the amplitudes of the latter {\it control channels}\/ are given by
\begin{equation}
    A(B_q^0 \to (c\bar c\, \bar dq)_h) = -\lambda {\cal A}'_h \left[ 1- e^{i\gamma} b'_h e^{i\theta'_h}\right],
\end{equation}
where in the limit of an exact flavour symmetry ${\cal A}'_h ={\cal A}_h$ and $b'_h e^{i\theta'_h}=b_h e^{i\theta_h}$.
Because the parameters $b'_h e^{i\theta'_h}$ are not suppressed by $\epsilon$, CP studies of the control modes are more sensitive to their extraction.
On the other hand, an overall Cabibbo rate suppression means their precision typically trails that of $b\to s c\bar c$ driven modes.

Proposed control channels for $B_s\to J/\psi \phi$ are the flavour symmetry related decays $B_d\to J/\psi \overline{K}^{0*}$ and $B_d\to J/\psi\rho^0$~\cite{Fleischer:1999zi,Faller:2008gt}.
As they too are vector-vector final states, each transversity amplitude can control the phase shift of its counterpart.
An important caveat is that $K^{0*}$ and $\rho^0$, as octets of $SU(3)_{\rm F}$, are related to the octet $\phi_8$, while the physical $\phi$, being mostly $s\bar s$, also has a sizable $SU(3)_{\rm F}$ singlet $\phi_0$ component~\cite{Gronau:2008hb}.
While $SU(3)_{\rm F}$ breaking is estimated to be $m_s/\Lambda_{\rm QCD}\sim f_{B_s}/f_{B_d}-1\sim 20{\rm -}30\%$, the reliability of a larger $U(3)$ nonet symmetry for these decays is less certain.

The mode $B_d\to J/\psi \overline{K}^{0*}$ is flavour-specific and thereby has no mixing-induced CP observables available.
Instead, its $C'_h$ observables must be complemented with decay rate information in order to extract $b'_h e^{i\theta'_h}$, introducing large uncertainties via the amplitude ratios $|{\cal A}_h/{\cal A}'_h|$~\cite{Faller:2008gt,Liu:2013nea}.
The branching ratio and polarization fractions of this decay have already been measured by LHCb~\cite{Aaij:2012nh}, and a 3~fb$^{-1}$ update that includes direct CP violation measurements is in progress\footnote{see the talk of W.~Kanso in these proceedings.}.
The mode $B_d\to J/\psi\rho^0$, on the other hand, does have access to mixing-induced CP observables and LHCb has recently reported a preliminary estimate of the penguin induced phase shift~\cite{Aaij:2014vda}.
Assuming exact $SU(3)_{\rm F}$ flavour symmetry, a universal phase shift of $\Delta\phi_f = (0.05\pm 0.56)^\circ$ is estimated.

The flavour symmetry method is also applied to control penguin effects in $B_d\to J/\psi K_{\rm S}$, where a similar phase shift is expected.
A recent update, including the control channels $B_d\to J/\psi \pi^0$ and $B_s\to J/\psi K_{\rm S}$, finds $\Delta\phi^{J/\psi K_{\rm S}} = (-0.97^{+0.72}_{-0.65})^\circ$~\cite{FleischerBEACH,DeBruyn:2010hh}, in agreement with earlier results~\cite{Faller:2008zc,Ciuchini:2011kd}.
In a similar study a full fit to the $SU(3)_{\rm F}$ related decays $B_{u,d,s}\to J/\psi\{K,\pi,(\eta_8)\}$ was performed, including linear breaking terms, finding $\Delta\phi^{J/\psi K_{\rm S}} \lesssim 1^\circ$~\cite{Jung:2012mp}.
In particular, the breaking terms were found to be crucial for the goodness of the fit.
A similar method could potentially be applied to the vector-vector decays $B_{u,d,s} \to J/\psi \{\phi,\omega,\rho,K^*\}$ in the future~\cite{Bediaga:2012py}.

There has recently been an attempt to calculate the non-perturbative contribution to the $u$-quark penguin in the low-energy effective theory~\cite{FringsCKM}.
By exploiting the large momentum flow through the $u$-quark loop (due to the heavy $J/\psi$) the penguin diagram can be shown to factorize into a single effective operator.
By placing bounds on the associated hadronic matrix element using a $1/N_c$ expansion, $|\Delta\phi_f| \leq 1.2^\circ$ is estimated.

\section{The nature of the $f_0(980)$ in $B$ decays}\label{sec:silver}

The $f_0(980)$ is a scalar state with the quantum numbers $J^{PC}=0^{++}$.
With a mass of 990~MeV it joins the club of scalar states with masses below 1~GeV that together could form a nonet, specifically $\{f_0(980),f_0(500),\kappa,a_0(980)$\}, the nature of which has been debated for several decades (for a review see Ref.~\cite{Klempt:2007cp}).
The simplest interpretation, that they are $q\bar q$ mesons in a P-wave, cannot explain their inverted mass hierarchy with respect to their assumed `strange' quark content.
Furthermore, the additional orbital momentum should make them half a GeV heavier than their vector meson nonet counterparts $\{\phi,\omega,K^*,\rho\}$.
Thus other interpretations that can explain these discrepancies have been suggested over the years.
These include tetraquarks (diquark--anti-diquark bound states), meson-meson molecules, or some mixture of the two. 

\begin{figure}[tb]
\centering
\raisebox{1.3cm}{\includegraphics[width=5cm]{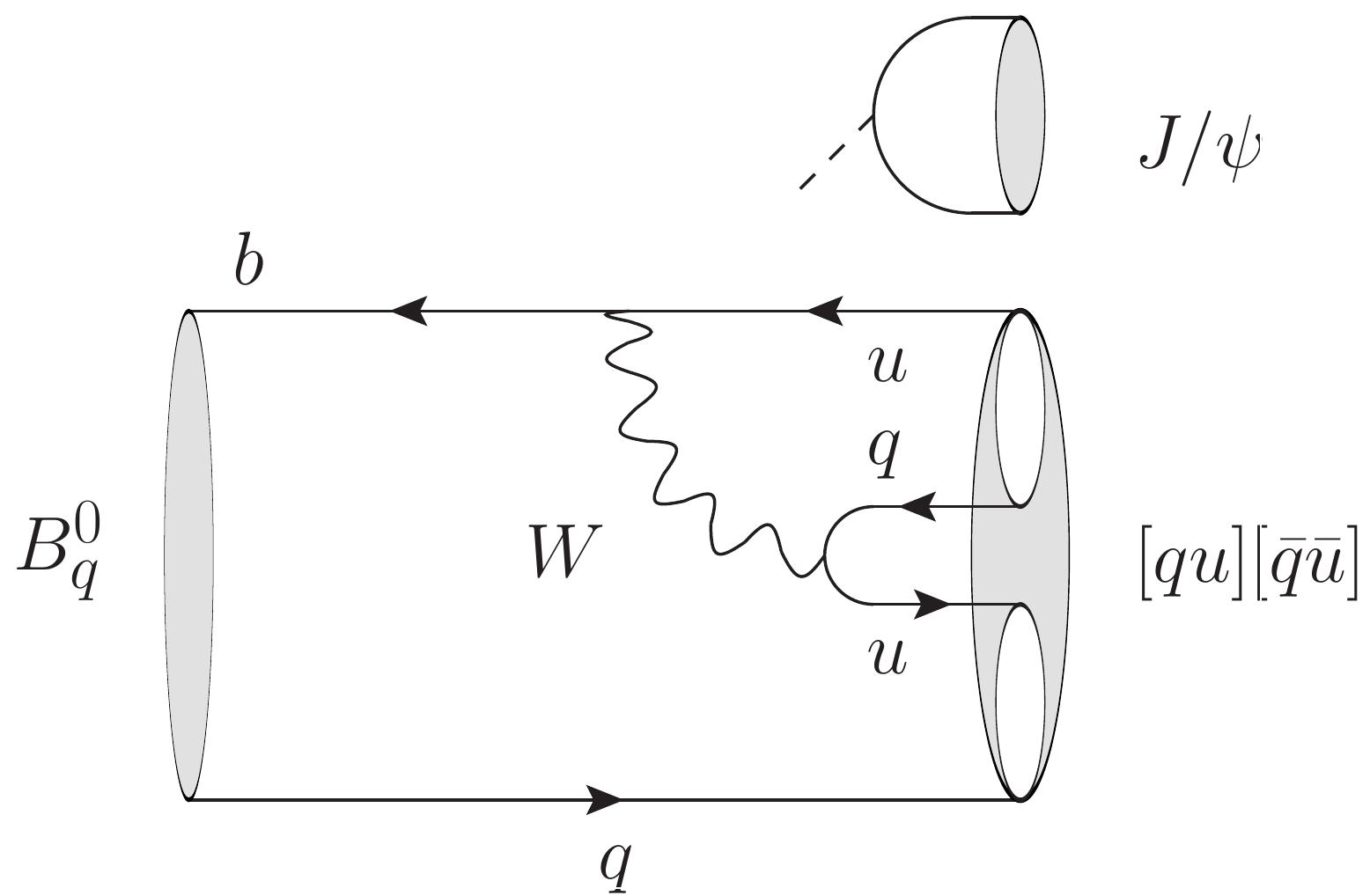}}
\hspace{1cm}
\includegraphics[width=7cm]{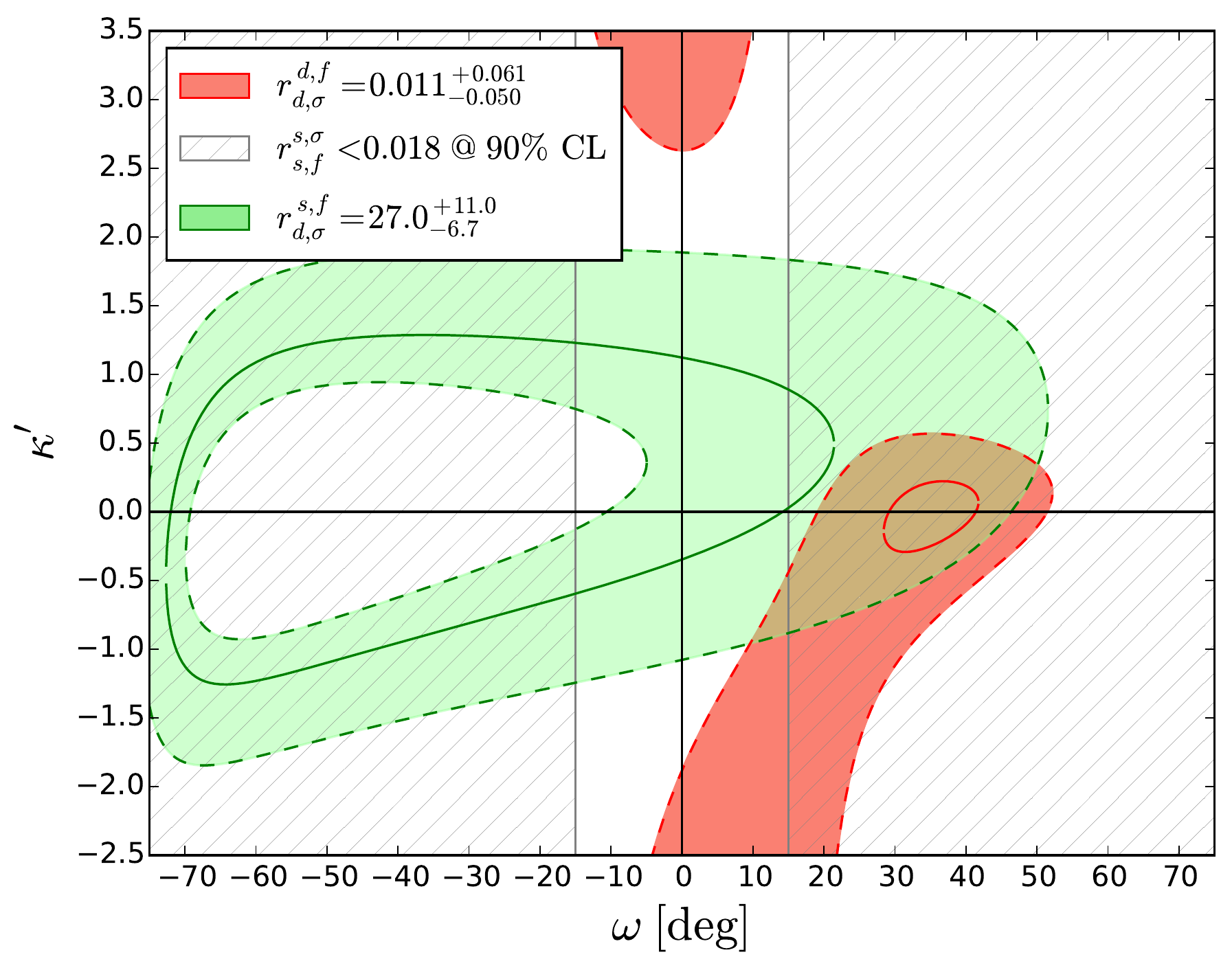}
\caption{{\it Left panel:}\/ the tetraquark diagram ${\cal A}^{(\prime)}_{4q}$ that contributes to $B_s^0 \to J/\psi\,f^{[0]}_0$ ($B_d^0 \to J/\psi\,\sigma^{[0]}$) for $q=s$ ($q=d$). 
{\it Right panel:}\/ constraints from $r^{d,f_0}_{d,\sigma}$, $r_{s,f_0}^{s,\sigma}$ and $r^{s,f}_{d,\sigma}$ in the tetraquark picture, with $\omega$ the mixing angle and $\kappa'$ the relative contribution of the ${\cal A}^{\prime}_{4q}$ diagram from the left panel (see text).}
\label{fig:kappaMix}
\end{figure}

In both the $q\bar q$ and tetraquark ($4q$) pictures the two isospin singlet states $f_0=f_0(980)$ and $\sigma=f_0(500)$ can mix.
We will consider simple realizations of both pictures, where the mixing is given by
\begin{equation}
\left(
\begin{array}{c}
f_0\\
\sigma
\end{array}
\right)\stackrel{q\bar{q}}{=}
\left(
\begin{array}{cc}
    \cos\varphi_{\rm M}\  & \ \sin\varphi_{\rm M} \\
-\sin\varphi_{\rm M}\  & \ \cos\varphi_{\rm M}
\end{array}
\right)
\cdot
\left(
\begin{array}{c}
s\bar s\\
\frac{1}{\sqrt{2}}\left(u \bar u + d\bar d\right)
\end{array}
\right),
\end{equation}
or
\begin{equation}
\left(
\begin{array}{c}
f_0\\
\sigma
\end{array}
\right)
\stackrel{4q}{=}
\left(
\begin{array}{cc}
\cos\omega\  & \ -\sin\omega \\
\sin\omega\  & \ \cos\omega
\end{array}
\right)
\cdot
\left(
\begin{array}{c}
f^{[0]}_0\\
\sigma^{[0]}
\end{array}
\right),\qquad
\left\{
\begin{array}{ccc}
f^{[0]}_0&\equiv&\frac{[su][\bar s \bar u]+[sd][\bar s\bar d]}{\sqrt{2}}\\
\sigma^{[0]}&\equiv&[ud][\bar u\bar d]
\end{array}
    \right..
\end{equation}
In Ref.~\cite{Fleischer:2011au} these two pictures were compared in the context of $B_s\to J/\psi f_0$, where it was found that significantly different decay dynamics are possible. 
In the tetraquark picture, in particular, an additional tree-level topology is present that carries a CP-violating phase, as shown in the left panel of Figure~\ref{fig:kappaMix}.
Based on such unknown decay dynamics the conservative estimate $\Delta\phi^{J/\psi f_0}\in [-3^\circ,3^\circ]$ was derived, and $B_d\to J/\psi f_0$ was explored as a potential control channel.
In Ref.~\cite{Stone:2013eaa} a test of the two pictures was proposed that utilises the mixing relation between the $f_0$ and the $\sigma$ via the branching ratio ratios~\cite{Stone:2013eaa,Li:2012sw}
\begin{align}
    r^{d,f_0}_{d,\sigma} \equiv \frac{{\rm BR}(B_d\to J/\psi\,f_0)}{{\rm BR}(B_d\to J/\psi\,\sigma)}\frac{\Phi_d(\sigma)}{\Phi_d(f_0)}
    &\stackrel{\dagger}{\sim} \left\{
        \begin{array}{ccc}
            \tan^2\varphi &:& q\bar q \\
            \frac{1}{2} &:& {\rm tetraquark}\ (\omega = 0^\circ)
        \end{array}
        \right.\label{rdd} \\
    r_{s,f_0}^{s,\sigma} \equiv \frac{{\rm BR}(B_s\to J/\psi\,\sigma)}{{\rm BR}(B_s\to J/\psi\,f_0)}\frac{\Phi_s(f_0)}{\Phi_s(\sigma)}
    &\stackrel{\dagger}{\sim} \left\{
        \begin{array}{ccc}
            \tan^2\varphi &:& q\bar q \\
            0 &:& {\rm tetraquark}\ (\omega = 0^\circ)
        \end{array}
        \right.,
        \label{rss}
\end{align}
where the choice of no tetraquark mixing was motivated by the bound $|\omega| \lesssim 5^\circ$ from Refs~\cite{Maiani:2004uc,Hooft:2008we,Fleischer:2011au}.
The above ratios have since been measured by LHCb, which finds $r^{d,f_0}_{d,\sigma} = 0.011^{+0.061}_{-0.050} < 0.098$~\cite{Aaij:2014siy} and $r_{s,f_0}^{s,\sigma} < 0.018$~\cite{Aaij:2014emv}, with the bounds at 90\%~CL.
If the relation given in \eqref{rdd} (denoted by $\dagger$) is taken to be exact, the tetraquark picture (without mixing) would be ruled out by $8\,\sigma$~\cite{Aaij:2014siy}.
If, in addition, the relation in \eqref{rss} were exact, the resulting small mixing angle would imply that the $f_0$ is almost completely a P-wave $s\bar s$ state in the $q\bar q$ picture.

There are, however, a number of caveats to the relations denoted by $\dagger$ in \eqref{rdd} and \eqref{rss} that need to be taken into account~\cite{KRF}.
For instance, there are sizable asymmetries possible in the production of the $f_0$ and $\sigma$ states.
If we were to consider, for example, only the leading colour-suppressed diagram and apply naive factorization, there is only limited evidence to suggest that $|F^{B_q \sigma}/F^{B_q f_0}|$ are close to one~\cite{Li:2012sw,Stone:2013eaa}.
There may also be significant sub-leading diagrams contributing, which for $b\to c\bar c d$ transitions receive no CKM suppressions.
In particular, the tetraquark topology shown in the left panel of Figure~\ref{fig:kappaMix} can contribute to $B_d\to J/\psi \sigma^{[0]}$ but not to $B_d\to J/\psi f^{[0]}_0$.
Furthermore, the assumption of negligible tetraquark mixing, namely $|\omega| \lesssim 5^\circ$ from Refs~\cite{Maiani:2004uc,Hooft:2008we}, could be outdated: updating this bound using a mass for $\kappa$ of $m_\kappa = 682$~MeV~\cite{Beringer:1900zz} (previously $m_\kappa = 797$~MeV), gives $\omega \approx 20^\circ$ (see also Ref.~\cite{Black:1998wt}).
Allowing for mixing modifies \eqref{rdd} and \eqref{rss} to
\begin{align}
    \left.r^{d,f_0}_{d,\sigma}\right|_{4q} \sim \frac{1}{2}\left|\frac{1-\sqrt{2}\tan\omega}{1+\frac{1}{\sqrt{2}}\tan\omega}\right|^2,\quad
    \left.r^{s,\sigma}_{s,f_0}\right|_{4q} \sim \left|\tan\left[\omega + \tan^{-1}\left(\sqrt{2} X_c\right)\right]\right|^2.\label{mixExpr}
\end{align}
Here $X_c$
is the ratio of exchange and penguin annihilation amplitudes relative to the dominant tree diagram. 
We estimate $|X_c|\lesssim 5\%$ from branching ratio limits on $B_d \to J/\psi \phi$, where these diagrams dominate, which is also the suppression of $\Lambda_{\rm QCD}/m_b$ expected for diagrams with spectator quark interactions.
An $X_c$ of this magnitude can shift $\omega$ in \eqref{mixExpr} by up to $\pm 5^\circ$.

Let $\kappa'$ denote the ratio of the special tetraquark topology for $B^0_d\to J/\psi \sigma^{[0]}$, shown in the left panel of Figure~\ref{fig:kappaMix}, with the leading $B^0_d\to J/\psi \sigma^{[0]}$ topologies\footnote{Our definition for $\kappa'$ is analogous to that of $b'$ in \eqref{AmplSS}, including also a factor $R_b$.}.
In the right panel of Figure~\ref{fig:kappaMix} we plot the experimental results for \eqref{rdd}, \eqref{rss}, and the ratio
\begin{align}
    r^{s,f}_{d,\sigma} \equiv
    {\frac{{\rm BR}(B_s\to J/\psi\, f_0)}{{\rm BR}(B_d\to J/\psi\, \sigma)}} 
    \frac{\Phi_d(\sigma)}{\Phi_s(f_0)},
\end{align}
with respect to the tetraquark mixing angle $\omega$ and $\kappa'$.
For simplicity we have ignored other sub-leading diagrams (such as penguins), and taken $\kappa'$ to be real.
Included in each amplitude ratio is also a 30\% error to account for the flavour symmetry breaking and possible production asymmetries for the scalar states.

We observe from the figure that by including these caveats the tetraquark picture can survive the current $B$ decay constraints, and it will be interesting to follow how this picture develops.
Of course not all of the discussed caveats necessarily apply, and Occam's razor currently favours a $q\bar q$ like production for the scalar states.
Either way, our descriptions for both the $q\bar q$ and tetraquark pictures are probably too simple, and the true nature of the light scalar states is unlikely to be solved by $B$ constraints alone.
Nonetheless, the important point is that a diagram like the one shown in the left panel of Figure~\ref{fig:kappaMix}, or its meson--meson molecular analogue (with a flipped internal quark line), could contribute significantly to the $B_s\to J/\psi \pi^+\pi^-$ decay modes, 
and thereby need to be controlled to allow for precise extractions of $\phi_s$.

\section{Summary}\label{sec:summary}

The $B_s$ mixing phase $\phi_s$ is now known to lie very close to the SM prediction thanks to excellent progress from the LHC experiments.
Therefore to continue the search for NP, the control of hadronic uncertainties in the $B_s\to J/\psi K^+ K^-$ and $B_d\to J/\psi \pi^+\pi^-$ modes will become a crucial aspect of these analyses.
To this end, we reviewed several available methods, including the control of penguin contributions using the flavour symmetry related modes $B_d^0 \to J/\psi K^{0*}$ and $B_d^0 \to J/\psi \rho$.
We also discussed the nature of the $f_0(980)$ resonance in the S-wave amplitudes of these modes, and argued that a tetraquark interpretation can survive the current constraints from $B$ decays.
As a result, the $f_0(980)$ resonance can exhibit different decay dynamics to that of the $\phi$, and care should be taken in interpreting averages of $\phi_s$ extractions involving the two.

\Acknowledgements
I would like to thank the conveners of working group IV for the invitation to this workshop, and the organizers for their organization and hospitality.
I am grateful to Giulia Ricciardi and Robert Fleischer for their collaboration on topics presented in this talk.
I would further like to thank Giulia and Martin Jung for their comments on this manuscript.

\bibliographystyle{h-physrev}
{\footnotesize\bibliography{references}}

\begin{thebibliography}{10}

\bibitem{Dighe:1995pd}
A.~S. Dighe, I.~Dunietz, H.~J. Lipkin, and J.~L. Rosner,
\newblock Phys.Lett. {\bf B369}, 144 (1996), hep-ph/9511363.

\bibitem{Dighe:1998vk}
A.~S. Dighe, I.~Dunietz, and R.~Fleischer,
\newblock Eur.Phys.J. {\bf C6}, 647 (1999), hep-ph/9804253.

\bibitem{Stone:2008ak}
S.~Stone and L.~Zhang,
\newblock Phys.Rev. {\bf D79}, 074024 (2009), 0812.2832.

\bibitem{Xie:2009fs}
Y.~Xie, P.~Clarke, G.~Cowan, and F.~Muheim,
\newblock JHEP {\bf 0909}, 074 (2009), 0908.3627.

\bibitem{Aaij:2013orb}
LHCb Collaboration, R.~Aaij {\em et~al.},
\newblock Phys.Rev. {\bf D87}, 072004 (2013), 1302.1213.

\bibitem{Stone:2009hd}
S.~Stone and L.~Zhang,
\newblock (2009), 0909.5442.

\bibitem{Aaij:2014emv}
LHCb Collaboration, R.~Aaij {\em et~al.},
\newblock Phys.Rev. {\bf D89}, 092006 (2014), 1402.6248.

\bibitem{Aaij:2014zsa}
R.~Aaij {\em et~al.},
\newblock (2014), 1411.3104.

\bibitem{Aad:2014cqa}
ATLAS Collaboration, G.~Aad {\em et~al.},
\newblock Phys.Rev. {\bf D90}, 052007 (2014), 1407.1796.

\bibitem{CMS-PAS-BPH-13-012}
CMS Collaboration,
\newblock CERN Report No. CMS-PAS-BPH-13-012, 2014 (unpublished).

\bibitem{Aaij:2014dka}
LHCb collaboration, R.~Aaij {\em et~al.},
\newblock Phys.Lett. {\bf B736}, 186 (2014), 1405.4140.

\bibitem{Charles:2004jd}
CKMfitter Group, J.~Charles {\em et~al.},
\newblock Eur. Phys. J. {\bf C41}, 1 (2005), hep-ph/0406184.

\bibitem{Faller:2008gt}
S.~Faller, R.~Fleischer, and T.~Mannel,
\newblock Phys.Rev. {\bf D79}, 014005 (2009), 0810.4248.

\bibitem{Boos:2004xp}
H.~Boos, T.~Mannel, and J.~Reuter,
\newblock Phys.Rev. {\bf D70}, 036006 (2004), hep-ph/0403085.

\bibitem{Li:2006vq}
H.-n. Li and S.~Mishima,
\newblock JHEP {\bf 0703}, 009 (2007), hep-ph/0610120.

\bibitem{Gronau:2008cc}
M.~Gronau and J.~L. Rosner,
\newblock Phys.Lett. {\bf B672}, 349 (2009), 0812.4796.

\bibitem{Fleischer:1999zi}
R.~Fleischer,
\newblock Phys.Rev. {\bf D60}, 073008 (1999), hep-ph/9903540.

\bibitem{Gronau:2008hb}
M.~Gronau and J.~L. Rosner,
\newblock Phys.Lett. {\bf B669}, 321 (2008), 0808.3761.

\bibitem{Liu:2013nea}
X.~Liu, W.~Wang, and Y.~Xie,
\newblock Phys.Rev. {\bf D89}, 094010 (2014), 1309.0313.

\bibitem{Aaij:2012nh}
LHCb Collaboration, R.~Aaij {\em et~al.},
\newblock Phys.Rev. {\bf D86}, 071102 (2012), 1208.0738.

\bibitem{Aaij:2014vda}
LHCb collaboration, R.~Aaij {\em et~al.},
\newblock (2014), 1411.1634.

\bibitem{FleischerBEACH}
K.~De~Bruyn and R.~Fleischer,
\newblock presented at BEACH 2014  (2014).

\bibitem{DeBruyn:2010hh}
K.~De~Bruyn, R.~Fleischer, and P.~Koppenburg,
\newblock Eur.Phys.J. {\bf C70}, 1025 (2010), 1010.0089.

\bibitem{Faller:2008zc}
S.~Faller, M.~Jung, R.~Fleischer, and T.~Mannel,
\newblock Phys.Rev. {\bf D79}, 014030 (2009), 0809.0842.

\bibitem{Ciuchini:2011kd}
M.~Ciuchini, M.~Pierini, and L.~Silvestrini,
\newblock (2011), 1102.0392.

\bibitem{Jung:2012mp}
M.~Jung,
\newblock Phys.Rev. {\bf D86}, 053008 (2012), 1206.2050.

\bibitem{Bediaga:2012py}
LHCb Collaboration, R.~Aaij {\em et~al.},
\newblock Eur.Phys.J. {\bf C73}, 2373 (2013), 1208.3355.

\bibitem{FringsCKM}
P.~Frings, U.~Nierste, and M.~Wiebusch,
\newblock presented at these proceedings  (2014).

\bibitem{Klempt:2007cp}
E.~Klempt and A.~Zaitsev,
\newblock Phys.Rept. {\bf 454}, 1 (2007), 0708.4016.

\bibitem{Fleischer:2011au}
R.~Fleischer, R.~Knegjens, and G.~Ricciardi,
\newblock Eur.Phys.J. {\bf C71}, 1832 (2011), 1109.1112.

\bibitem{Stone:2013eaa}
S.~Stone and L.~Zhang,
\newblock Phys.Rev.Lett. {\bf 111}, 062001 (2013), 1305.6554.

\bibitem{Li:2012sw}
J.-W. Li, D.-S. Du, and C.-D. Lu,
\newblock Eur.Phys.J. {\bf C72}, 2229 (2012), 1212.5987.

\bibitem{Maiani:2004uc}
L.~Maiani, F.~Piccinini, A.~Polosa, and V.~Riquer,
\newblock Phys.Rev.Lett. {\bf 93}, 212002 (2004), hep-ph/0407017.

\bibitem{Hooft:2008we}
G.~'t~Hooft, G.~Isidori, L.~Maiani, A.~Polosa, and V.~Riquer,
\newblock Phys.Lett. {\bf B662}, 424 (2008), 0801.2288.

\bibitem{Aaij:2014siy}
LHCb Collaboration, R.~Aaij {\em et~al.},
\newblock Phys.Rev. {\bf D90}, 012003 (2014), 1404.5673.

\bibitem{KRF}
R.~Fleischer, G.~Ricciardi, and R.~Knegjens,
\newblock work in progress  (2014).

\bibitem{Beringer:1900zz}
Particle Data Group, J.~Beringer {\em et~al.},
\newblock Phys.Rev. {\bf D86}, 010001 (2012).

\bibitem{Black:1998wt}
D.~Black, A.~H. Fariborz, F.~Sannino, and J.~Schechter,
\newblock Phys.Rev. {\bf D59}, 074026 (1999), hep-ph/9808415.

\end{thebibliography}
 
\end{document}